\documentclass{svjour3}
\usepackage{graphicx}
\usepackage{wasysym}

\begin{document}

\title{Relation between various formulations of perturbation equations of
celestial mechanics}

\author{Pavol P\'{a}stor}

\institute{P. P\'{a}stor \at
   Department of Astronomy, Physics of the Earth, and Meteorology, \\
   Faculty of Mathematics, Physics and Informatics, \\
   Comenius University,
   Mlynsk\'{a} dolina, 842~48 Bratislava, Slovak Republic \\
   \email{pavol.pastor@fmph.uniba.sk}
}

\date{}

\authorrunning{P. P\'{a}stor}
\titlerunning{Implication between perturbation equations of celestial
mechanics}

\maketitle

\begin{abstract}
Orbital motion of a body can be found from Newtonian equation of motion.
However, it is useful to express the motion through time derivatives of
Keplerian orbital elements, mainly if the motion is perturbed by small
perturbing force. The first set of equations for the time derivatives
of the orbital elements can be derived from the equation of motion using
Lagrange brackets. The second one by using equation of motion
and perturbation acceleration decomposed to radial, transversal and normal
components. This paper shows that the second type of the perturbation
equations can be derived from the first type using simple mathematical
operations.
\end{abstract}

\section{Introduction}

Perturbation equations of celestial mechanics belong to the standard part of
the classical astronomy. Perturbation acceleration represents disturbing
force which disturbs Keplerian motion. Disturbing force causes time change
of orbital elements describing actual orbit of a body in space. To describe
the orbit of the body in space we can use different sets of orbital elements.
We will use the following set of orbital elements: semi-major axis $a$,
eccentricity $e$, inclination $i$, argument of perihelion $\omega$,
longitude of ascending node $\Omega$ and angle $\sigma=n \tau$, where $n$ is
mean motion and $\tau$ is time of perihelion passage. Perturbation equations
can be expressed using different methods. Using Lagrange method it is possible
to derive perturbation equations which use scalar product of disturbing
acceleration and partial derivative of position vector with respect to
orbital elements. Another method is to derive perturbation equations
directly from radial, transversal and normal components of disturbing force.
We show that the expression obtained by Lagrange method enables to
derive the expression through radial, transversal and normal
components of the disturbing force. Attempts to show the existence of this
connection can be found in Brown (1896). Brown uses an alternate set
of orbital elements and his equations contain several trivial errors.

\section{Expression obtained using Lagrange brackets}

Using Lagrange brackets, we can derive the following time derivatives of
orbital elements defined in previous section (see, e. g., Brouwer
and Clemence 1961)
\begin{eqnarray}\label{1-6}
\frac{da}{dt} &=& \frac{2}{na} ~\vec{a}_{D} \cdot
      \frac{ \partial \vec{r}}{ \partial \sigma} ~, \\
\frac{de}{dt} &=& \frac{1-e^{2}}{na^{2}e} ~\vec{a}_{D} \cdot
      \frac{ \partial \vec{r}}{ \partial \sigma} -
      \frac{ \sqrt{1-e^{2}}}{na^{2}e} ~\vec{a}_{D} \cdot
      \frac{ \partial \vec{r}}{ \partial \omega} ~, \\
\frac{di}{dt} &=& \frac{ \cot i}{na^{2} \sqrt{1-e^{2}}} ~\vec{a}_{D} \cdot
      \frac{ \partial \vec{r}}{ \partial \omega} -
      \frac{1}{na^{2} \sqrt{1-e^{2}} \sin i} ~\vec{a}_{D} \cdot
      \frac{ \partial \vec{r}}{ \partial \Omega} ~, \\
\frac{d \sigma}{dt} &=& - \frac{2}{na} ~\vec{a}_{D} \cdot
      \frac{ \partial \vec{r}}{ \partial a} -
      \frac{1-e^{2}}{na^{2}e} ~\vec{a}_{D} \cdot
      \frac{ \partial \vec{r}}{ \partial e} ~, \\
\frac{d \omega}{dt} &=& \frac{\sqrt{1-e^{2}}}{na^{2}e} ~\vec{a}_{D} \cdot
      \frac{ \partial \vec{r}}{ \partial e} -
      \frac{ \cot i}{na^{2} \sqrt{1-e^{2}}} ~\vec{a}_{D} \cdot
      \frac{ \partial \vec{r}}{ \partial i} ~, \\
\frac{d \Omega}{dt} &=& \frac{1}{na^{2} \sqrt{1-e^{2}} \sin i} ~
      \vec{a}_{D} \cdot \frac{ \partial \vec{r}}{ \partial i} ~,
\end{eqnarray}
where $\vec{a}_{D}$ is a disturbing acceleration and $\vec{r}$ is a position
vector of a particle with respect to the Sun.

\section{Expression through radial, transversal and normal component of
disturbing acceleration}
We can express time derivatives of orbital elements though radial, transversal
and normal components of disturbing acceleration in the following way
\begin{eqnarray}\label{7-12}
\frac{da}{dt} &=& \frac{2}{n \sqrt{1-e^{2}}}
      \left [a_{R} ~e \sin f + a_{T} ~(1+e \cos f) \right ] ~, \\
\frac{de}{dt} &=& \frac{ \sqrt{1-e^{2}}}{na}
      \left [a_{R} ~\sin f + a_{T} \left ( \cos f +
      \frac{e + \cos f}{1+e \cos f} \right ) \right ] ~, \\
\frac{di}{dt} &=& a_{N} ~\frac{r \cos \Theta}{na^{2} \sqrt{1-e^{2}}} ~, \\
\frac{d \sigma}{dt} &=& \frac{1-e^{2}}{na} \left [a_{R} \left (
      \frac{ \cos f}{e} - \frac{2}{1+e \cos f} \right ) - a_{T} ~
      \frac{ \sin f}{e} \frac{2+e \cos{f}}{1+e \cos f} \right ] -
      t \frac{dn}{dt} ~, \\
\frac{d \omega}{dt} &=& \frac{ \sqrt{1-e^{2}}}{nae} \left (-a_{R} ~\cos f +
      a_{T} ~\sin f \frac{2+e \cos f}{1+e \cos f} \right ) - a_{N} ~
      \frac{r \sin \Theta \cot i}{na^{2} \sqrt{1-e^{2}}} ~, \\
\frac{d \Omega}{dt} &=& a_{N} ~
      \frac{r \sin \Theta}{na^{2} \sqrt{1-e^{2}} \sin i} ~,
\end{eqnarray}
where $r=a(1-e^{2})/(1+e \cos f)$ for an elliptical orbit and
$\Theta= f + \omega$. See e. g. Bate et al. (1971), Kla\v{c}ka (1992).

\section{Relation}
When we want to evaluate time derivatives of orbital elements in Eqs. (1)-(6),
we need to know the values of
$\vec{a}_{D} \cdot ( \partial \vec{r} / \partial A)$, where
$A \in \{ a, e, i, \sigma, \omega, \Omega \}$. Moreover, we want to
express time derivatives in Eqs. (1)-(6) through radial, transversal and normal
components of perturbation acceleration $\vec{a}_{D}$ $=$ $a_{R} \vec{e}_{R} +
a_{T} \vec{e}_{T} + a_{N} \vec{e}_{N}$. To do this, we express also components
of the vector $\vec{u}_{A} = \partial \vec{r} / \partial A$ through radial,
transversal and normal components. We have
\begin{eqnarray}\label{13}
\vec{u}_{A} &=& \frac{ \partial \vec{r}}{ \partial A} =
      \frac{ \partial x}{ \partial A} \vec{i} +
      \frac{ \partial y}{ \partial A} \vec{j} +
      \frac{ \partial z}{ \partial A} \vec{k} =
      u_{Ax} \vec{i} +
      u_{Ay} \vec{j} +
      u_{Az} \vec{k} =
\nonumber \\
&=&   u_{AR} ~\vec{e}_{R} +
      u_{AT} ~\vec{e}_{T} +
      u_{AN} ~\vec{e}_{N} ~,
\end{eqnarray}
where $\vec{i}$, $\vec{j}$, $\vec{k}$ are unit vectors in directions of
coordinate axes $x$, $y$, $z$ of Cartesian coordinate system. Since vectors
$\vec{e}_{R}$, $\vec{e}_{T}$, $\vec{e}_{N}$ are orthonormal, we can write for
radial, transversal and normal components of the vector $u_{A}$
\begin{eqnarray}\label{14-16}
u_{AR} &=& u_{Ax} \vec{i} \cdot \vec{e}_{R} +
      u_{Ay} \vec{j} \cdot \vec{e}_{R} +
      u_{Az} \vec{k} \cdot \vec{e}_{R} =
      \vec{u}_{A} \cdot \vec{e}_{R} ~, \\
u_{AT} &=& u_{Ax} \vec{i} \cdot \vec{e}_{T} +
      u_{Ay} \vec{j} \cdot \vec{e}_{T} +
      u_{Az} \vec{k} \cdot \vec{e}_{T} =
      \vec{u}_{A} \cdot \vec{e}_{T} ~, \\
u_{AN} &=& u_{Ax} \vec{i} \cdot \vec{e}_{N} +
      u_{Ay} \vec{j} \cdot \vec{e}_{N} +
      u_{Az} \vec{k} \cdot \vec{e}_{N} =
      \vec{u}_{A} \cdot \vec{e}_{N} ~.
\end{eqnarray}
Finally, we get for $\vec{a}_{D} \cdot ( \partial \vec{r} / \partial A)$
\begin{equation}\label{17}
\vec{a}_{D} \cdot \frac{ \partial \vec{r}}{ \partial A} =
a_{R} ~u_{AR} + a_{T} ~u_{AT} + a_{N} ~u_{AN} ~.
\end{equation}

\begin{figure}
\begin{center}
\includegraphics{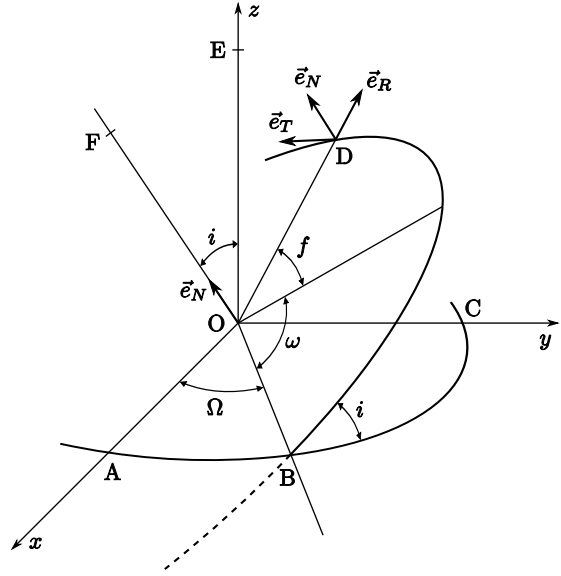}
\end{center}
\caption{Oscular orbital elements and radial, transversal and normal unit
vectors.}
\end{figure}

We need to find components of unit vectors $\vec{e}_{R}$, $\vec{e}_{T}$,
$\vec{e}_{N}$ in Cartesian coordinate system as a function of orbital elements.
We can write
\begin{equation}\label{18}
\vec{e}_{R} = (\cos \alpha_{1}, \cos \alpha_{2}, \cos \alpha_{3}) ~,
\end{equation}
\begin{equation}\label{19}
\vec{e}_{T} = (\cos \beta_{1}, \cos \beta_{2}, \cos \beta_{3}) ~,
\end{equation}
\begin{equation}\label{20}
\vec{e}_{N} = (\cos \gamma_{1}, \cos \gamma_{2}, \cos \gamma_{3}) ~,
\end{equation}
where $\alpha_{1}$, $\alpha_{2}$ and $\alpha_{3}$ are angles between vector
$\vec{e}_{R}$ and coordinate axes $x$, $y$ and $z$, respectively. Similarly for
vectors $\vec{e}_{T}$ and $\vec{e}_{N}$. To calculate one of the vectors
$\vec{e}_{R}$, $\vec{e}_{T}$, $\vec{e}_{N}$, we can use cross product of others
two, since vectors are orthonormal. Components of the vectors can be found
using a spherical law of cosines
\begin{equation}\label{21}
\cos \angle{\rm POS} = \cos \angle{\rm POR} ~\cos \angle{\rm ROS} +
      \sin \angle{\rm POR} ~\sin \angle{\rm ROS} ~\cos \delta ~,
\end{equation}
where each of the points P, R, S lie on one of the three different lines
crossing in the point O (the point O is different from the points P, R, S)
and $\delta$ is the angle between planes determined by points P, O, R and
R, O, S.

We can find components of the unit vector $\vec{e}_R$ from spherical triangles
$\triangle$ABD, $\triangle$DBC and $\triangle$EBD which can be constructed in
Fig. 1. We will use notation $\Theta=f+\omega$. From spherical
triangle $\triangle$ABD we have
\vspace{0.3cm}\\
$\angle$BOA $= \Omega$, $\angle$DOB $= \Theta$,
$\delta_{\alpha_{1}} =\pi - i$,\\
$\cos \alpha_{1} = \cos \angle$AOD$ = \cos \angle$BOA$ ~\cos \angle$DOB$ +
\sin \angle$BOA$ ~\sin \angle$DOB$ ~\cos \delta_{\alpha_{1}}$,\\
$\cos \alpha_{1} = \cos \Omega \cos \Theta - \sin \Omega \sin \Theta \cos i$,
\vspace{0.3cm}\\
where $\delta_{\alpha_{1}}$ is angle between planes determined by points
B, O, A and D, O, B. From spherical triangle $\triangle$DBC
\vspace{0.3cm}\\
$\angle$DOB $= \Theta$, $\angle$COB $= \pi/2 - \Omega$,
$\delta_{\alpha_{2}} =i$,\\
$\cos \alpha_{2} = \cos \angle$DOC$ = \cos \angle$DOB$ ~\cos \angle$COB$ +
\sin \angle$DOB$ ~\sin \angle$COB$ ~\cos \delta_{\alpha_{2}}$,\\
$\cos \alpha_{2} = \cos \Theta \sin \Omega + \sin \Theta \cos \Omega \cos i$,
\vspace{0.3cm}\\
where $\delta_{\alpha_{2}}$ is the angle between the planes determined by
points D, O, B and C, O, B. From spherical triangle $\triangle$EBD
we obtain
\vspace{0.3cm}\\
$\angle$EOB $= \pi/2$, $\angle$DOB $= \Theta$,
$\delta_{\alpha_{3}} =\pi/2 - i$,\\
$\cos \alpha_{3} = \cos \angle$EOD$ = \cos \angle$EOB$ ~\cos \angle$DOB$ +
\sin \angle$EOB$ ~\sin \angle$DOB$ ~\cos \delta_{\alpha_{3}}$,\\
$\cos \alpha_{3} = \sin \Theta \sin i$,
\vspace{0.3cm}\\
where $\delta_{\alpha_{3}}$ is the angle between the planes determined by
points E, O, B and D, O, B.

Components of unit vector $\vec{e}_N$ can be found from spherical triangles
$\triangle$ABF, $\triangle$BCF and from the angle $\angle$FOE depicted in
Fig. 1. From spherical triangle $\triangle$ABF we have
\vspace{0.3cm}\\
$\angle$BOA $= \Omega$, $\angle$BOF $= \pi/2$,
$\delta_{\gamma_{1}} =\pi/2 - i$,\\
$\cos \gamma_{1} = \cos \angle$AOE$ = \cos \angle$BOA$ ~\cos \angle$BOF$ +
\sin \angle$BOA$ ~\sin \angle$BOF$ ~\cos \delta_{\gamma_{1}}$,\\
$\cos \gamma_{1} = \sin \Omega \sin i$,
\vspace{0.3cm}\\
where $\delta_{\gamma_{1}}$ is the angle between the planes determined by
points B, O, A and B, O, F. From spherical triangle $\triangle$CFB we have
\vspace{0.3cm}\\
$\angle$BOF $= \pi/2$, $\angle$COB $= \pi/2 - \Omega$,
$\delta_{\gamma_{2}} =\pi/2 + i$,\\
$\cos \gamma_{2} = \cos \angle$FOC$ = \cos \angle$BOF$ ~\cos \angle$COB$ +
\sin \angle$BOF$ ~\sin \angle$COB$ ~\cos \delta_{\gamma_{2}}$,\\
$\cos \gamma_{2} = - \cos \Omega \sin i$,
\vspace{0.3cm}\\
where $\delta_{\gamma_{2}}$ is the angle between the planes determined by
the points B, O, A and B, O, F. For $\angle$FOE we have
\vspace{0.3cm}\\
$\cos \gamma_{3} = - \cos \angle$FOE $ = \cos i$.
\vspace{0.3cm}

Components of unit vector $\vec{e}_{T}$ we can calculate using cross product
$\vec{e}_{N} \times \vec{e}_{R}$. We can summarize the results as
\begin{eqnarray}\label{22-24}
\vec{e}_{R} &=& \left (\cos \Omega \cos (f+ \omega) -
      \sin \Omega \sin (f+ \omega) \cos i, \right.
\nonumber \\
& &   \left. \sin \Omega \cos (f+ \omega) +
      \cos \Omega \sin (f+ \omega) \cos i,~
      \sin (f+ \omega) \sin i \right ) ~,\\
\vec{e}_{T} &=& \left ( - \cos \Omega \sin (f+ \omega) -
      \sin \Omega \cos (f+ \omega) \cos i, \right.
\nonumber \\
& &   \left. - \sin \Omega \sin (f+ \omega) +
      \cos \Omega \cos (f+ \omega) \cos i,~
      \cos (f+ \omega) \sin i \right ) ~,\\
\vec{e}_{N} &=& (\sin \Omega \sin i, - \cos \Omega \sin i, \cos i) ~,
\end{eqnarray}
where we have used $\Theta=f+ \omega$ again.

We begin with evaluating of $da/dt$. We need to calculate partial derivatives
$\partial x/ \partial \sigma$, $\partial y/ \partial \sigma$ and
$\partial z/ \partial \sigma$ in Eq. (1). We can find components of the
position vector from the relation $\vec{r}=r ~\vec{e}_{R}$, where
\begin{equation}\label{25}
r=\frac{a(1-e^{2})}{1+e \cos f} ~,
\end{equation}
for the elliptical orbit. Finally, we get for $x$, $y$, $z$
\begin{equation}\label{26}
x = \frac{ \big [ \cos \Omega \cos (f+ \omega) -
      \sin \Omega \sin (f+ \omega) \cos i \big ] ~a(1-e^{2})}{1+e \cos f} ~,
\end{equation}
\begin{equation}\label{27}
y = \frac{ \big [ \sin \Omega \cos (f+ \omega) +
      \cos \Omega \sin (f+ \omega) \cos i \big ] ~a(1-e^{2})}{1+e \cos f} ~,
\end{equation}
\begin{equation}\label{28}
z = \frac{ \sin (f+ \omega) \sin i ~a(1-e^{2})}{1+e \cos f} ~.
\end{equation}
In these expressions only $f$ is a function of $\sigma$, thus we need to
calculate $\partial f/\partial \sigma$. For this purpose we use Kepler
equation in the form
\begin{equation}\label{29}
nt+\sigma=E-e \sin E ~,
\end{equation}
where E is eccentric anomaly. We calculate partial derivative of
Eq. (29) with respect to $\sigma$ and the result rewrite to the form
\begin{equation}\label{30}
\frac{ \partial E}{ \partial \sigma}=\frac{1}{1-e \cos E} ~.
\end{equation}
Now we use relation between the true anomaly and eccentric anomaly
\begin{equation}\label{31}
f = 2 \arctan \left ( \sqrt { \frac{1+e}{1-e}} \tan \frac{E}{2} \right ) ~.
\end{equation}
Partial derivative of Eq. (31) with respect to $\sigma$ give
\begin{equation}\label{32}
\frac{ \partial f}{ \partial \sigma} = \frac{1}{ \sqrt{1-e^{2}}} ~
(1+e \cos f) ~\frac{ \partial E}{ \partial \sigma} ~,
\end{equation}
where we have used also the relation
\begin{equation}\label{33}
\cos E = \frac{e + \cos f}{1 + e \cos f} ~.
\end{equation}
We put Eq. (30) into Eq. (32) and the result we rewrite to the following form
\begin{equation}\label{34}
\frac{ \partial f}{ \partial \sigma} = \frac{a^{2} \sqrt{1-e^{2}}}{r^{2}} ~.
\end{equation}
Now we can calculate partial derivatives
$\partial x/ \partial \sigma$, $\partial y/ \partial \sigma$ and
$\partial z/ \partial \sigma$
\begin{eqnarray}\label{35}
\frac{ \partial x}{ \partial \sigma} &=&u_{ \sigma x} =
      \frac{a}{ \sqrt{1-e^{2}}} ~
      \big \{- \cos \Omega ~[ \sin (f + \omega) + e \sin \omega]
\nonumber \\
& &   - \sin \Omega \cos i ~[ \cos (f + \omega) + e \cos \omega] \big \} ~,
\end{eqnarray}
\begin{eqnarray}\label{36}
\frac{ \partial y}{ \partial \sigma} &=& u_{ \sigma y} =
      \frac{a}{ \sqrt{1-e^{2}}} ~
      \big \{- \sin \Omega ~[ \sin (f + \omega) + e \sin \omega]
\nonumber \\
& &   + \cos \Omega \cos i ~[ \cos (f + \omega) + e \cos \omega] \big \} ~,
\end{eqnarray}
\begin{eqnarray}\label{37}
\frac{ \partial z}{ \partial \sigma} &=& u_{ \sigma z} =
      \frac{a}{ \sqrt{1-e^{2}}} ~
      \sin i ~[ \cos (f + \omega) + e \cos \omega] ~.
\end{eqnarray}
When we put equations Eqs. (35)-(37) and Eq. (22) into Eq. (14), we get for
the radial component of the vector $\vec{u}_{ \sigma}$ the following relation
\begin{equation}\label{38}
u_{ \sigma R} = \frac{a}{ \sqrt{1-e^{2}}} ~e \sin f ~.
\end{equation}
Similarly, from Eq. (15) using Eqs. (35)-(37) and Eq. (23) we get for
the transversal component
\begin{equation}\label{39}
u_{ \sigma T} = \frac{a}{ \sqrt{1-e^{2}}} ~(1 + e \cos f) ~,
\end{equation}
and, finally, for the normal component, we get, from Eq. (16) using
Eqs. (35)-(37) and Eq. (24),
\begin{equation}\label{40}
u_{ \sigma N} = 0 ~.
\end{equation}
Using Eqs. (38)-(40) in Eq. (17) we can now calculate
$\vec{a}_{D} \cdot ( \partial \vec{r} / \partial \sigma)$
\begin{equation}\label{41}
\vec{a}_{D} \cdot \frac{ \partial \vec{r}}{ \partial \sigma} =
      \frac{a}{ \sqrt{1-e^{2}}} ~[a_{R} ~e \sin f + a_{T} ~(1 + e \cos f)] ~.
\end{equation}
Putting Eq. (41) into Eq. (1) we obtain for time derivative of the semimajor
axis
\begin{equation}\label{42}
\frac{da}{dt} = \frac{2}{n \sqrt{1-e^{2}}} ~
      [a_{R} ~e \sin f + a_{T} ~(1 + e \cos f)] ~.
\end{equation}
This is the relation identical to Eq. (7).

Now we want to calculate time derivative of eccentricity. We need to evaluate
$\vec{a}_{D} \cdot ( \partial \vec{r} / \partial \omega)$ in Eq. (2).
For partial derivatives $\partial x/ \partial \omega$,
$\partial y/ \partial \omega$, $\partial y/ \partial \omega$ we obtain, from
Eqs. (26)-(28),
\begin{equation}\label{43}
\frac{ \partial x}{ \partial \omega} = u_{ \omega x} =
      \frac{ \big [ - \cos \Omega \sin (f+ \omega) -
      \sin \Omega \cos (f+ \omega) \cos i \big ] ~a(1-e^{2})}{1+e \cos f} ~,
\end{equation}
\begin{equation}\label{44}
\frac{ \partial y}{ \partial \omega} = u_{ \omega y} =
      \frac{ \big [ - \sin \Omega \sin (f+ \omega) +
      \cos \Omega \cos (f+ \omega) \cos i \big ] ~a(1-e^{2})}{1+e \cos f} ~,
\end{equation}
\begin{equation}\label{45}
\frac{ \partial z}{ \partial \omega} = u_{ \omega z} =
      \frac{ \cos (f+ \omega) \sin i ~a(1-e^{2})}{1+e \cos f} ~.
\end{equation}
This equations can be more simply written in our notation as (see also Eq. 23)
\begin{equation}\label{46}
\frac{ \partial \vec{r}}{ \partial \omega} = \vec{u}_{ \omega} =
      \frac{a(1-e^{2})}{1+e \cos f} ~\vec{e}_{T} = r ~\vec{e}_{T} ~.
\end{equation}
Using Eq. (46) and Eqs. (14)-(16) we can immediately write
\begin{equation}\label{47}
u_{ \omega R} = 0 ~,
\end{equation}
\begin{equation}\label{48}
u_{ \omega T} = r ~,
\end{equation}
\begin{equation}\label{49}
u_{ \omega N} = 0 ~.
\end{equation}
By inserting Eqs. (47)-(49) into Eq. (17) we obtain
\begin{equation}\label{50}
\vec{a}_{D} \cdot \frac{ \partial \vec{r}}{ \partial \omega} = a_{T} ~r ~.
\end{equation}
Inserting Eq. (41) and Eq. (50) into Eq. (2) we obtain for the time derivative
of the eccentricity
\begin{equation}\label{51}
\frac{de}{dt} = \frac{ \sqrt{1-e^{2}}}{na}
      \left [a_{R} ~\sin f + a_{T} \left ( \cos f +
      \frac{e + \cos f}{1+e \cos f} \right ) \right ] ~.
\end{equation}
This relation is identical with Eq. (8).

Next we want to calculate $di/dt$. In this order we need to evaluate
$\vec{a}_{D} \cdot ( \partial \vec{r} / \partial \Omega)$ in
Eq. (3). For partial derivatives of coordinates with respect to $\Omega$ we
get
\begin{equation}\label{52}
\frac{ \partial x}{ \partial \Omega} = u_{ \Omega x} =
      \frac{ \big [ - \sin \Omega \cos (f+ \omega) -
      \cos \Omega \sin (f+ \omega) \cos i \big ] ~a(1-e^{2})}{1+e \cos f} ~,
\end{equation}
\begin{equation}\label{53}
\frac{ \partial y}{ \partial \Omega} = u_{ \Omega y} =
      \frac{ \big [ \cos \Omega \cos (f+ \omega) -
      \sin \Omega \sin (f+ \omega) \cos i \big ] ~a(1-e^{2})}{1+e \cos f} ~,
\end{equation}
\begin{equation}\label{54}
\frac{ \partial z}{ \partial \Omega} = u_{ \Omega z} = 0 ~.
\end{equation}
For radial, transversal and normal components of the vector $\vec{u}_{ \Omega}$
we obtain
\begin{equation}\label{55}
u_{ \Omega R} = 0 ~,
\end{equation}
\begin{equation}\label{56}
u_{ \Omega T} = r \cos i ~,
\end{equation}
\begin{equation}\label{57}
u_{ \Omega N} = - r \cos (f+ \omega) \sin i ~.
\end{equation}
By inserting Eqs. (55)-(57) into Eq. (17) we obtain
\begin{equation}\label{58}
\vec{a}_{D} \cdot \frac{ \partial \vec{r}}{ \partial \Omega} =
      r [a_{T} ~\cos i - a_{N} ~\cos (f+ \omega) \sin i] ~.
\end{equation}
From Eq. (3), using Eq. (50) and Eq. (58), we get
\begin{equation}\label{59}
\frac{di}{dt} = a_{N} ~\frac{r \cos (f+ \omega)}{na^{2} \sqrt{1-e^{2}}} ~.
\end{equation}
This relation is identical with Eq. (9).

In order to calculate $d \sigma / dt$, we need to know
$\vec{a}_{D} \cdot \partial \vec{r} / \partial a$ and
$\vec{a}_{D} \cdot \partial \vec{r} / \partial e$ in Eq. (4).
We begin with $\vec{a}_{D} \cdot \partial \vec{r} / \partial a$. In
Eqs. (26)-(28) we must take into account that also $f$ is a function of $a$.
Thus, we need to calculate $\partial f/ \partial a$. We can use Eq. (31) to
find similarly as in Eq. (32)
\begin{equation}\label{60}
\frac{ \partial f}{ \partial a} = \frac{1}{ \sqrt{1-e^{2}}} ~
(1+e \cos f) ~\frac{ \partial E}{ \partial a} ~.
\end{equation}
Now we use Kepler equation defined in Eq. (29). Partial derivative of Eq. (29)
with respect to $a$ gives the following relation
\begin{equation}\label{61}
\frac{ \partial E}{ \partial a} = \frac{1}{1-e \cos E}
      \frac{ \partial n}{ \partial a} t =
      \frac{1}{1-e \cos E} \frac{dn}{da}t ~,
\end{equation}
where $n$ is the mean motion $n=\sqrt{GM/a^{3}}$ ($G$ is the gravitational
constant and $M$ is mass of central object). Using Eq. (34)
(compare also Eqs. 32 and 30 with Eqs. 60 and 61) we can write for
$\partial f / \partial a$
\begin{equation}\label{62}
\frac{ \partial f}{ \partial a} = \frac{ \partial f}{ \partial \sigma}
      \frac{dn}{da}t ~.
\end{equation}
Using Eqs. (35)-(37) we can write for partial derivatives of coordinates $x$,
$y$, $z$ with respect to $a$
\begin{equation}\label{63}
\frac{ \partial x}{ \partial a} = u_{ a x} =
      \frac{ \big [ \cos \Omega \cos (f+ \omega) -
      \sin \Omega \sin (f+ \omega) \cos i \big ](1-e^{2})}{1+e \cos f} +
      \frac{ \partial x}{ \partial \sigma} \frac{dn}{da} t ~,
\end{equation}
\begin{equation}\label{64}
\frac{ \partial y}{ \partial a} = u_{ a y} =
      \frac{ \big [ \sin \Omega \cos (f+ \omega) +
      \cos \Omega \sin (f+ \omega) \cos i \big ](1-e^{2})}{1+e \cos f} +
      \frac{ \partial y}{ \partial \sigma} \frac{dn}{da} t ~,
\end{equation}
\begin{equation}\label{65}
\frac{ \partial z}{ \partial a} = u_{ a z} =
      \frac{ \sin (f+ \omega) \sin i ~(1-e^{2})}{1+e \cos f} +
      \frac{ \partial z}{ \partial \sigma} \frac{dn}{da} t ~.
\end{equation}
These three equations can be more simple written in our notation as
\begin{equation}\label{66}
\frac{ \partial \vec{r}}{ \partial a} = \vec{u}_{a} =
      \frac{(1-e^{2})}{1+e \cos f} ~\vec{e}_{R} +
      \frac{dn}{da} t ~\vec{u}_{ \sigma} ~.
\end{equation}
By inserting Eq. (66) into Eqs. (14)-(16) we obtain
\begin{equation}\label{67}
u_{a R} = \frac{(1-e^{2})}{1+e \cos f} + u_{ \sigma R} ~\frac{dn}{da} t ~,
\end{equation}
\begin{equation}\label{68}
u_{a T} = u_{ \sigma T} ~\frac{dn}{da} t ~,
\end{equation}
\begin{equation}\label{69}
u_{a N} = u_{ \sigma N} ~\frac{dn}{da} t ~.
\end{equation}
When we put Eqs. (67)-(69) into Eq. (17) we get
\begin{equation}\label{70}
\vec{a}_{D} \cdot \frac{ \partial \vec{r}}{ \partial a} =
      a_{R} ~\frac{(1-e^{2})}{1+e \cos f} +
      \vec{a}_{D} \cdot \frac{ \partial \vec{r}}{ \partial \sigma}
      \frac{dn}{da} t ~.
\end{equation}
We can now use Eq. (1) to obtain
\begin{equation}\label{71}
\vec{a}_{D} \cdot \frac{ \partial \vec{r}}{ \partial a} =
      a_{R} ~\frac{(1-e^{2})}{1+e \cos f} +
      \frac{na}{2} \frac{da}{dt} \frac{dn}{da} t =
      a_{R} ~\frac{(1-e^{2})}{1+e \cos f} + \frac{na}{2} \frac{dn}{dt} t ~.
\end{equation}
Moreover, we need to evaluate
$\vec{a}_{D} \cdot \partial \vec{r} / \partial e$ in order to find $de/dt$.
In Eqs. (26)-(28) we need take into account that $f$ is also a function of $e$.
For partial derivative of Kepler equation defined by Eq. (29) with respect to
$e$ we obtain
\begin{equation}\label{72}
\frac{ \partial E}{ \partial e} = \frac{ \sin E}{1 - e \cos E} ~.
\end{equation}
When we use Eq. (72) and also expressions $a (1-e \cos E)$ $=$ $r$ and
$a \sqrt{1-e^{2}} \sin E$ $=$ $r \sin f$ in partial derivative of Eq. (31)
with respect to $e$ we finally obtain
\begin{equation}\label{73}
\frac{ \partial f}{ \partial e} = \frac{\sin f}{1-e^{2}}(2+e \cos f) ~.
\end{equation}
For partial derivatives of coordinates $x$, $y$, $z$ we can,
in our notation, write
\begin{equation}\label{74}
\frac{ \partial \vec{r}}{ \partial e} = r \frac{ \partial f}{ \partial e} ~
      \vec{e}_{T} - \left [ \frac{2ae}{1+e \cos f} +
      \frac{a(1-e^{2})}{(1+e \cos f)^{2}}
      \left ( \cos f - e \sin f \frac{ \partial f}{ \partial e} \right )
      \right ] \vec{e}_{R} ~,
\end{equation}
This equation can be simplified using Eq. (73) as
\begin{equation}\label{75}
\frac{ \partial \vec{r}}{ \partial e} = \vec{u}_{e} = a \sin f ~
      \frac{2+e \cos f}{1+e \cos f} ~\vec{e}_{T} - a \cos f ~\vec{e}_{R} ~.
\end{equation}
By inserting Eq. (75) into Eqs. (14)-(16) we obtain
\begin{equation}\label{76}
u_{e R} = - a \cos f ~,
\end{equation}
\begin{equation}\label{77}
u_{e T} = a \sin f ~\frac{2+e \cos f}{1+e \cos f} ~,
\end{equation}
\begin{equation}\label{78}
u_{e N} = 0 ~.
\end{equation}
When we put Eqs. (76)-(78) into Eq. (17) we get
\begin{equation}\label{79}
\vec{a}_{D} \cdot \frac{ \partial \vec{r}}{ \partial e} = -
      a_{R} ~a \cos f + a_{T} ~a \sin f ~\frac{2+e \cos f}{1+e \cos f} ~.
\end{equation}
Now we can use Eq. (71) and Eq. (79) in Eq. (4) to calculate $d \sigma /dt$
we have
\begin{equation}\label{80}
\frac{d \sigma}{dt} = \frac{1-e^{2}}{na} \left [a_{R} \left (
      \frac{ \cos f}{e} - \frac{2}{1+e \cos f} \right ) - a_{T} ~
      \frac{ \sin f}{e} \frac{2+e \cos{f}}{1+e \cos f} \right ] -
      t \frac{dn}{dt} ~.
\end{equation}
This relation is identical with Eq. (10).

Now we want to calculate time derivative of argument of perihelion. In this
order we need to evaluate
$\vec{a}_{D} \cdot ( \partial \vec{r} / \partial i)$ in
Eq. (5). For partial derivatives $\partial x/ \partial i$,
$\partial y/ \partial i$, $\partial y/ \partial i$ we obtain from
Eqs. (26)-(28)
\begin{equation}\label{81}
\frac{ \partial x}{ \partial i} = u_{ i x} =
      \frac{ \sin \Omega \sin (f+ \omega) \sin i ~a(1-e^{2})}
      {1+e \cos f} ~,
\end{equation}
\begin{equation}\label{82}
\frac{ \partial y}{ \partial i} = u_{ i y} =
      \frac{- \cos \Omega \sin (f+ \omega) \sin i ~a(1-e^{2})}
      {1+e \cos f} ~,
\end{equation}
\begin{equation}\label{83}
\frac{ \partial z}{ \partial i} = u_{ i z} =
      \frac{ \sin (f+ \omega) \cos i ~a(1-e^{2})}{1+e \cos f} ~.
\end{equation}
For radial, transversal and normal components of vector $\vec{u}_{i}$ we obtain
from Eqs. (14)-(16)
\begin{equation}\label{84}
u_{ i R} = 0 ~,
\end{equation}
\begin{equation}\label{85}
u_{ i T} = 0 ~,
\end{equation}
\begin{equation}\label{86}
u_{ i N} = r \sin (f+ \omega) ~.
\end{equation}
From Eq. (17), using Eqs. (84)-(86), we get
\begin{equation}\label{87}
\vec{a}_{D} \cdot \frac{ \partial \vec{r}}{ \partial i} = a_{N} ~
      r \sin (f+ \omega) ~.
\end{equation}
When we now put Eqs. (79) and (87) into Eq. (5), we obtain
\begin{equation}\label{88}
\frac{d \omega}{dt} = \frac{ \sqrt{1-e^{2}}}{nae} \left (-a_{R} ~\cos f +
      a_{T} ~\sin f \frac{2+e \cos f}{1+e \cos f} \right ) - a_{N} ~
      \frac{r \sin (f+ \omega) \cot i}{na^{2} \sqrt{1-e^{2}}} ~, \\
\end{equation}
This relation is identical with Eq. (11).

Now we can finally put Eq. (87) into Eg. (6), in order to obtain
$d \Omega / dt$
\begin{equation}\label{89}
\frac{d \Omega}{dt} = a_{N} ~
      \frac{r \sin (f+ \omega)}{na^{2} \sqrt{1-e^{2}} \sin i} ~.
\end{equation}
This relation is identical with Eq. (12).

\section{Conclusion}

We have just shown that it is possible to derive Eqs. (7)-(12) from
Eqs. (1)-(6) using partial derivatives of position vector with respect
to orbital elements.

\begin{acknowledgements}
The paper was supported by the Scientific Grant Agency VEGA
(grant No. 2/0016/09).
\end{acknowledgements}

\end{document}